\documentclass[twocolumn]{jpsj3} %% two-column layout
\usepackage{bm} %bold math

\newcommand{\degree}{{$^\circ$}}

\title{Space efficient opposed-anvil high-pressure cell and its application to
optical and NMR measurements up to 9 GPa}

\author{Kentaro~\textsc{Kitagawa$^1$}\thanks{kitag@issp.u-tokyo.ac.jp},
Naoyuki~\textsc{Katayama$^1$}\thanks{present address: Department of Physics,
University of Virginia, Charlottesville, Virginia, USA},
Kenya~\textsc{Ohgushi$^{1,2}$}, and Masashi~\textsc{Takigawa$^{1,2}$}}

\author{Kentaro~\textsc{Kitagawa$^1$}\thanks{kitag@issp.u-tokyo.ac.jp},
Hirotada~\textsc{Gotou$^1$},
Takehiko~\textsc{Yagi$^1$},
Atsushi~\textsc{Yamada$^{2,1}$},
Takehiko~\textsc{Matsumoto$^1$},
Yoshiya~\textsc{Uwatoko$^1$},
Masashi~\textsc{Takigawa$^1$}}

\inst{$^1$Institute for Solid State Physics, University of Tokyo,
5-1-5 Kashiwanoha, Kashiwa, Chiba 277-8581, Japan\\
$^2$Graduate School of Science and Engineering, Saitama Univ., Saitama,
Saitama 338-8570, Japan}

\abst{We have developed a new type of opposed-anvil high pressure cell with 
substantially improved space efficiency. The clamp cell and the gasket are made 
of non-magnetic Ni-Cr-Al alloy. Non-magnetic tungsten carbide
(NMWC) is used for the anvils. The assembled cell with the dimension $\phi
29$~mm $\times 41$~mm is capable of generating pressure up to 9~GPa over a relatively large volume of  
7~mm$^3$.  Our cell is particularly suitable for those experiments which require 
large sample space to achieve good signal-to-noise ratio, such as the nuclear 
magnetic resonance (NMR) experiment. Argon is used as the pressure transmitting 
medium to obtain good hydrostaticity. The pressure was calibrated in situ by 
measuring the fluorescence from ruby through a transparent moissanite (6H-SiC) 
window. We have measured the pressure and temperature dependences of the 
$^{63}$Cu nuclear-quadrupole-resonance (NQR) frequency of Cu$_2$O, 
the in-plane Knight shift of metallic tin, and the Knight shift of platinum. These 
quantities can be used as reliable manometers to determine the pressure values 
in situ during the NMR/NQR experiments up to 9~GPa.}
\kword{high pressure, opposed-type anvil, Ni-Cr-Al alloy, NMR, NQR, cuprous,
tin, platinum, ruby fluorescence, argon}

\begin{document}
\maketitle

\section{Introduction}
Application of high pressure is a common method in solid
state physics.  A reduction of lattice constant under pressure modifies lattice
or electronic structure of matters and opens possibility to discover new states 
of matters.  At the same time, apparatus to generate high pressure generally 
impose experimental constraints due to limited volume of the pressurized 
space.  An easier and frequently used alternative is to utilize the "chemical pressure" 
by substituting some elements with other chemically similar elements with 
different atomic sizes. However, chemical substitutions inevitably introduce 
structural disorder and electronic inhomogeneity, which often change
the properties and the phase diagram of the pure materials.  
Therefore, it is important to develop space efficient high pressure apparatus 
to make various experimental techniques accessible to high pressure.

Generally, one has to compromise between demands for larger space and 
higher pressure.  For example, there is an order-of-thousand difference 
in the available sample space between the piston-cylinder-type 
cell\cite{UwatokoNiCrAl}, which can be used to generate pressure less 
than 4~GPa, and the modified Bridgeman-type cell\cite{FukazawaBridgeman}, 
which can go up to 10~GPa. Thus, easy access to the pressure above 4~GPa 
is practically restricted to volume-insensitive experiments such as resistivity
or optical measurements.  On the other hand, recent studies on strongly correlated 
electron systems containing transition metal elements with $d$-electrons
revealed interesting quantum phase transitions taking place in the pressure range above
4~GPa. The examples include $\beta$-phase vanadate
bronzes\cite{YamauchiNaBlonzeSC} and  iron pnictide
superconductors\cite{Alireza122HP,Matsubayashi122HP,KitagawaSrFe2As2UHP}.
For microscopic understanding of these phenomena, various experiments including
 volume-sensitive measurements such as nuclear magnetic resonance (NMR) are required.
Thus there is a increasing demand for high-pressure cells 
with improved space-efficiency.   

Our aim is to develop a high pressure cell which enables realistic nuclear 
magnetic resonance (NMR) experiments well above 4~GPa.
Although a few trials have been reported using modified
Bridgeman\cite{FukazawaBridgeman,SuzukiHP} cells, moissanite anvil
cell (MAC)\cite{HaasNMR}, or diamond anvil cells
(DAC)\cite{PravicaDACNMR,OkuchiDACNMR,OkuchiDACNMR2}, 
problems still remain concerning the insufficient sample space,
poor hydrostaticity, or inhomogeneity of the magnetic
field due to magnetization of the cells\cite{OkuchiDACNMR,SuzukiHP}. In this paper, we report development of a
new type of opposed-anvil-cell which can be used to generate pressure up to 9~GPa over a volume 
of 7~mm$^3$ with the total cell size of $\phi 29$~mm  $\times 41$~mm. This
sample space is ten times larger than that reported for the
modified-Bridgeman-type cell\cite{FukazawaBridgeman,SuzukiHP} and hundreds
times than MAC\cite{HaasNMR} or
DAC\cite{PravicaDACNMR,OkuchiDACNMR,OkuchiDACNMR2}. Our cell has three additional advantages.
First, a transparent moissanite window can be attached for optical
measurements, which enables accurate determination of pressure by measuring the fluorescence from ruby.
Second, the compact size of the cell allows arbitrary rotation of the whole cell by a two-axis goniometer to achieve precise 
alignment of single crystal samples in superconducting magnets.
Third, argon can be employed as pressure transmitting medium.
Argon is soft molecular solid at high pressures and provides a highly
hydrostatic environment\cite{MaoRubyAr,TateiwaMedia}.
This feature is particularly important at low temperatures, where
any liquid medium inevitably solidifies\cite{TateiwaMedia}. 

Since the optical setup to measure ruby fluorescence occupies one third of
the sample space and the cell cannot be rotated with the optical fiber
attached, it is necessary to establish NMR pressure indicators.  For this purpose, we have
measured the
precise pressure ($P$)- and temperature ($T$)- dependences of the nuclear 
quadrupole resonance (NQR) frequency of $^{63}$Cu nuclei in cuprous oxide 
(Cu$_{2}$O) and the Knight shift of metallic copper, $\beta$-tin, and platinum 
between 4.2--300~K up to 9~GPa against the ruby pressure scale\cite{MaoRubyAr,JayaramanReview}.

The NQR frequencies ($^{63}\nu_{Q}$) of Cu$_2$O is known to show a significant
pressure dependence\cite{ReyesCu2O}.
Reyes \textit{et al.} have established Cu$_2$O-NQR as a suitable manometer in
the temperature range 4--300~K and for the pressure up to 2~GPa\cite{ReyesCu2O}. Fukazawa \textit{et al.}
have extended the measurements up to 10~GPa at 1.6~K\cite{FukazawaBridgeman}.
We found that the $T$- and $P$-dependences of $^{63}\nu_{Q}$ can be precisely fit to a 
formula, which is a refined version of the formula used by Wijin and Wildt\cite{WijnCu2O} 
and by Reyes~\textit{et\,al.}\cite{ReyesCu2O} based upon a simple phononic analysis.
The Knight shift of $^{63}$Cu shows a very small $P$-dependence, $-10$~ppm/GPa,    
therefore it can be used to determine the field values in the pressure cell during 
NMR experiments. To the contrary, the shifts of $^{119}$Sn and $^{195}$Pt have
significant $P$-dependence: $-150, 220$~ppm/GPa respectively in the low-$H,T$
limit. These results are consistent with the previous studies below 1~GPa at
room temperatures.\cite{MatzkaninSnPbPt,BenedekPdep} The Sn/Pt NMR has less
accuracy as a manometer than the NQR frequency of Cu$_2$O.  However they are useful 
in magnetic field where the broad NMR spectrum from Cu$_2$O may overlap with 
the signal one wants to measure.    
\begin{figure}[htb!p]
\centering
\includegraphics[width=0.9\linewidth]{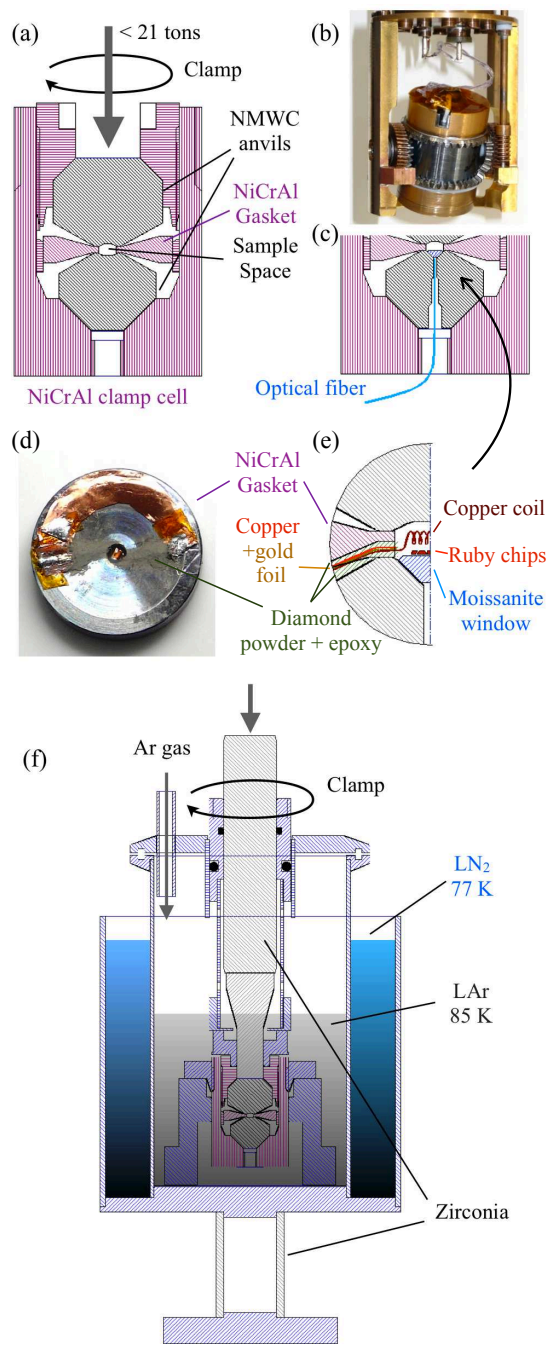}
\caption{(Color online) The cross-sectional view of the high-pressure clamp cell 
without (a) and with (c) the optical setup. Two anvils made of non-magnetic tungsten 
carbide (NMWC) press the gasket and sample space. For optical measurements, 
a moissanite cone is attached to the face of one of the anvils. (b) The whole cell can be 
mounted on an geared two-axis goniometric stage for arbitrary alignment of the sample 
in magnetic field. (d) A pickup coil and wires settled on a gasket.
(e) The enlarged view for the setup (c). (f) The cross-sectional view of the
argon loading system. Liquid nitrogen in the outer pot liquifies argon.
The load is transmitted through the rod made by alumina-zirconia composite ceramics, which has low thermal conductivity. }
\label{fig:schem}
\end{figure}

\section{Experiment}
\subsection{Non-magnetic clamp cell: anvils, gasket, and wiring}
Figure~\ref{fig:schem} shows the drawings of the pressure cell. The two facing
anvils are made of non-magnetic WC (NMWC) alloy (Fujilloy MF10). The anvil
without a window has a tapered cavity of 3~mm in diameter and 0.35~mm in depth. 
The top face of the anvils is 4~mm in diameter and surrounded by a conical face 
with a lateral slope angle of 30\degree. For the optical setup [Fig.~\ref{fig:schem}(c)], 
moissanite (single-crystal
6H-SiC, manufactured by Charles \& Colvard) of 2.5~mm in diameter is attached in
a conical hollow of an NMWC anvil and glued with epoxy resin and
diamond powder. The gasket and clamp cell are made of non-magnetic 56Ni-40Cr-4Al alloy, which has the 
tensile strength of 2~GPa.\cite{UwatokoNiCrAl} NiCrAl alloy of gaskets and the
clamp cell is age-hardened at 740 and 800\degree C for 12 hours respectively to
obtain good balances between tensile strength and toughness. We have examined
about forty different shapes and materials for the gasket to maximize the volume between the 
anvils over 8~GPa. The best performance was obtained by an anticonical-shaped 
gasket with appropriate apertures to the anvils. While the common shape of 
gaskets for a opposed-type-anvil cell is plain disk with a hole, we found
the additional slope can augment the gasket and wiring against shear
force from pressure transmitting medium. The slope angle of the gasket is chosen 
to be 15\degree, the slope begins at 2.35~mm off the center, and the hole is 2.5~mm 
in diameter and 1~mm in thickness. Combined with the cavities of the anvils, 
the total initial sample space is about 7~mm$^3$. 
Wiring was settled in grooves with 1~mm wide and 0.66~mm depth milled on the
gasket [Figs.~\ref{fig:schem}(d) and (e)]. The cooper wire from
the pickup coil plated with gold and copper foil were insulated by diamond powder with epoxy resin.

\subsection{Low-temperature loading of argon,
pressure transmitting medium}
Argon is used as the pressure transmitting
medium unless otherwise noted, by loading the cell in liquid argon [Fig.~\ref{fig:schem}(f)].
Argon, which liquifies at 87.3~K at 1~atm, is cooled by liquid nitrogen.
To press the entire cell at low
temperatures, the piston and support are made of zirconia with low thermal conductivity.
During the loading process, temperature of
the liquid argon is stabilized by heating at 85~K, slightly above the melting
 point of Ar at 1~atm (83.8~K). After loading typically up to 7~tons the cell
is warmed back to room temperature and the argon is sealed by clamping the
cell. We found that sealing of argon has to be performed above 3~GPa in order
to prevent leak of Ar gas if it is to be kept at room temperature for a long
period. To further increase the pressure, additional load is applied at room
temperature up to 21~tons through the WC rod with 10~mm diameter.
Hydraulic pressures have been smoothly changed by an electrically
controlled plunger pump.

\subsection{NMR/NQR experiments and ruby fluorescence}
We have used the following reagents in the present study: 
Cu$_{2}$O powder (99.9\%, Rare metallic),
platinum powder (99.98\%, NILACO) and tin powder (99.999\%, NEWMET KOCH).
To observe signals of copper metal, we mainly used
pickup coils of standard copper wires.
To check possible effects of purity, elongation, or residual strains, we
examined the difference at ambient pressure between powder and other forms: the reagent
powder and single crystals of cuprous oxide grown according to
Ref.~\citen{NakanoCu2OSC}; copper of coil wire and 99.9\% of copper powder
(SOEKAWA); the reagent foil and reagent powder for platinum and tin.
No difference in the NQR frequencies or in the Knight shifts (in-plane
components for tin) were observed within the experimental resolution determined
by the line width.
All the measurements except for copper in the following sections have been
performed in powder form. 

NMR experiment has
been performed using conventional pulsed spectrometers.
The spectra were acquired as Fourier transform of the spin echo signal. We regard the
peak positions as the representative values. The Knight shifts measurements of
NMR for platinum and tin have been performed by comparing the
NMR frequencies $^{195}$Pt or $^{119}$Sn with $^{63}$Cu at the same
field around at 6.6~T. For the Knight shifts of copper, we compared
$^{63}$Cu NMR with proton NMR of glycerol at the same field around 4.2~T. 

\begin{figure}[htbp]
\centering
\includegraphics[width=1.0\linewidth]{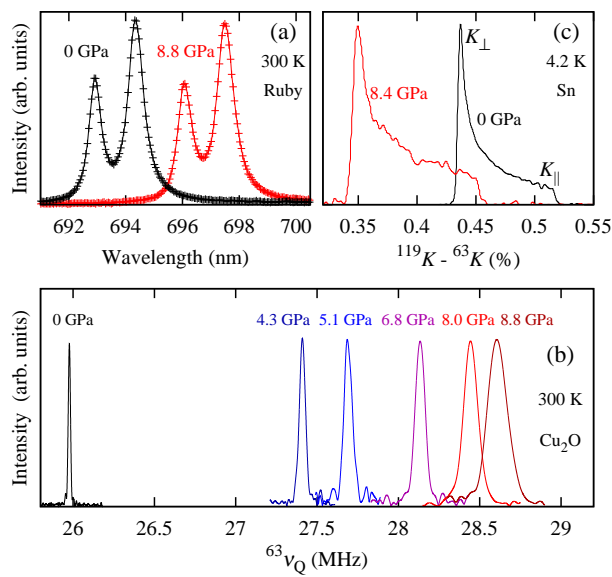}
\caption{(Color online) Spectra of manometers at various 
pressures in argon medium.
(a) Ruby fluorescence spectra at ambient pressure and at 8.8~GPa at 300~K.
(b) $^{63}$Cu-NQR spectra of Cu$_2$O at 300~K. 
(c) $^{119}$Sn-NMR spectra of metallic tin at 4.2~K. Since the $\beta$-Sn has a 
tetragonal crystal structure, the NMR spectrum is broadened by uniaxial anisotropy.  
The peak represents the Knight shift for the field in the $ab$-plane.}
\label{fig:spectra}
\end{figure}

Pressure values are determined by the ruby fluorescence method. It is based on the nearly
linear $P$-dependence of the shift of the wavelength $\lambda$ for the R1
fluorescence peak: $d\lambda/dP = 0.365$~nm/GPa in the low-$P$ region\cite{PiermariniRuby}. 
This relation is known to be insensitive to temperature. In this study, we used a more precise
non-linear equation reported by Mao~\textit{et.\,al.}, who used argon medium up to 
80~GPa:\cite{MaoRubyAr}
\begin{equation}
P = \frac{1904}{7.665}\left\{\left(1 +  \frac{
\lambda - \lambda_0}{\lambda_0}\right)^{7.665} - 1\right\}\text{  (GPa)}.
\end{equation}
Here $\lambda_0$ is the R1 wavelength at ambient pressure and at the same
temperature. We measured $\lambda$ and $\lambda_0$ inside and outside the 
pressurized space using two pieces of ruby from the same batch. Typical fluorescence spectra 
at room temperature are shown in Fig.~\ref{fig:spectra}(a).  

The NMR/NQR measurements were performed as follows.  First the cell was loaded 
at room temperature until the desired pressure was reached.  Then the cell was clamped
to hold the pressure, mounted on the NMR probe, and installed in the He
cryostat. The NMR/NQR measurements were performed at several stabilized temperatures 
between 300 and 4.2~K.  The pressure changes to some extent during the cooling process 
because the elasticity of Ni-Cr-Al alloy is strengthened and
molecular motion of medium slows at lower-$T$. To determine the
pressure accurately at each temperature, we measured ruby fluorescence at zero field just before and/or after the 
NMR/NQR measurements. For low $P$ and high $T$ region, argon slowly leaks 
as detected by decrease of $P$.  Therefore we performed the NMR/NQR measurements  
only in high-$P$ or low-$T$ region where the pressure
determination is reliable.

\section{Application to NMR/NQR}
\subsection{NQR frequency of $\text{Cu}_2$O}
Cu$_{2}$O is a semiconductor, in which the Cu-O bonds have ionic nature.
The moderate relaxation times and the sharp NQR transition line in Cu$_{2}$O 
make this material a suitable pressure indicator.  
Figure~\ref{fig:spectra}(b) displays the NQR spectra of $^{63}$Cu nuclei at
300~K at several pressures. The spectrum at ambient pressure is very sharp 
with the full width at half maximum (FWHM) of 20~kHz centered at 25.98~MHz. 
The NQR frequency  $^{63}\nu_\text{Q}$ is proportional to the electric field 
gradient (EFG) at the nuclei, which is axially symmetric around the O-Cu-O bond.  
Generally EFG from a distant point charge has the $ r^{-3}$ dependence. Therefore,
$\nu_\text{Q}$ is expected to be inversely proportional to the volume. In fact, $^{63}\nu_\text{Q}$ 
in Cu$_{2}$O increases as $P$ is  increased as shown in
Fig.~\ref{fig:spectra}(b) and Table~\ref{tab:cu2o}. As  previously reported,
$^{63}\nu_\text{Q}$ shows almost linear dependence on $P$ with the slope of 0.33~MHz/GPa at 300~K. 
Larger FWHM at higher $P$ is considered to be due to inhomogeneity of pressure 
and/or axial stress in powder sample. Therefore the FWHM is a good measure of 
the hydrostaticity.  The FWHM of 100~kHz at 8~GPa in our opposite-anvil cell 
with Ar medium is smaller than the previous result of 140~kHz
obtained with Flourinert medium and cubic-anvil apparatus\cite{HirayamaNQR}.
This ensures good hydrostaticity of our system.

\begin{figure}[htbp]
\centering
\includegraphics[width=1.0\linewidth]{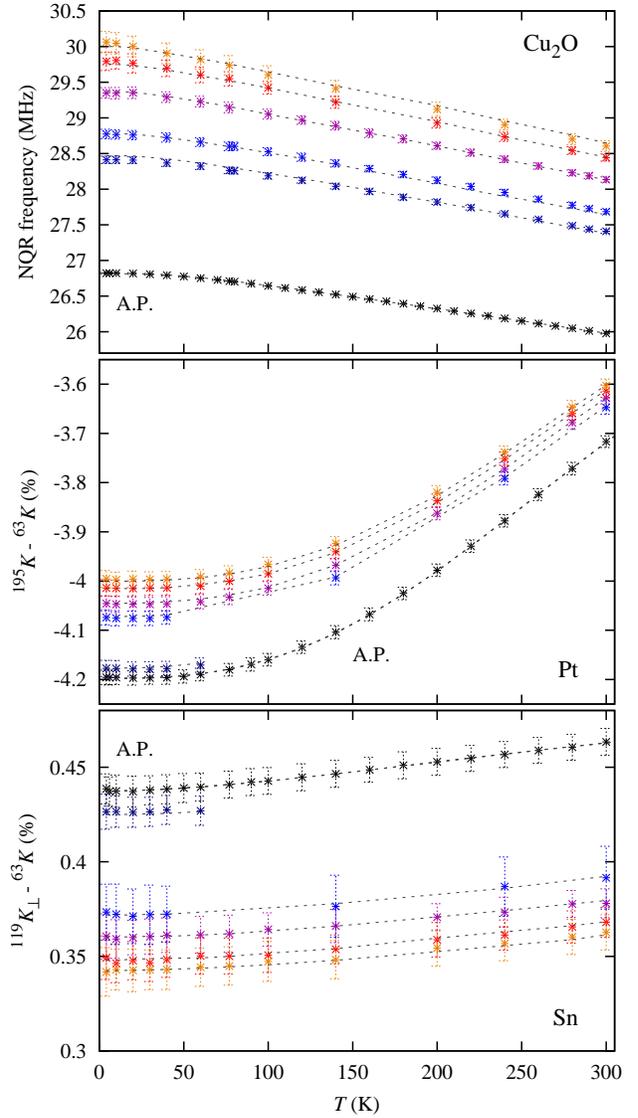}
\caption{(Color online) The temperature dependence of NMR/NQR manometers in
several runs of clamping. For each data point, pressure is determined by the ruby
scale. For the clamped pressures, see the Fig.~4 and 5, where these 
quantities are plotted against $P$. The dashed lines connect the fitted results at each $(P, T)$ (see text
for the formulas). The errorbars represent the (equivalent) FWHM.}
\label{fig:vst}
\end{figure}

Reyes~\textit{et\,al.} presented an analytical formula\cite{ReyesCu2O} to express 
$^{63}\nu_\text{Q}$ as a function of $P$ up to 2~GPa and $T$ between 4 and 300~K,
by combining the $T$-dependence due to thermally activated phonons \cite{WijnCu2O} 
and the linear $P$-dependence, 
\begin{align}\label{eq:cu2o}
&^{63}\nu_\text{Q}(P, T)\\ 
&= {^{63}\nu_\text{Q}^{(0)}(P)}\left\{1 +
\lambda(P)\Theta(P)\left(\frac{\exp(\frac{\Theta(P)}{T})}{1 -
\exp(\frac{\Theta(P)}{T})} + \frac{1}{2}\right)\right\}\\ &=
{^{63}\nu_\text{Q}^{(0)}(P)}\left\{1 - \frac{\lambda(P)\Theta(P)}{2}\coth\left(\frac{\Theta(P)}{2T}\right)\right\}.
\end{align}
Here $^{63}\nu_\text{Q}^{(0)}(P)$ is the $T$-independent coefficient, $\Theta$
is the characteristic temperature for the relevant phonon mode and $\lambda$
is the coupling constant between the phonon and EFG. 
These parameters should be a function of volume $V$ or of $P$.  Below we generalize 
the treatment by Reyes~\textit{et\,al.} by allowing non-linear $P$-dependence 
of these parameters.  In Fig.~\ref{fig:vst}, $^{63}\nu_\text{Q}$ is
plotted against $T$ at ambient pressure and at several pressures obtained by clamping
the cell.  The above equation provides excellent fit to the experimental points 
at ambient pressure (dashed line) with a standard deviation of 4~kHz,
which is much less than the FWHM of 20~kHz:
\begin{align}
{^{63}\nu_\text{Q}^{(0)}(P = 0)} &= 27.0513(50)~\text{ MHz},\\
\lambda(P=0) &= 0.00012978(65)~\text{ K$^{-1}$},\\
\Theta(P=0) &= 135.3(27)~\text{ K}.
\end{align}

\begin{table}[h]
\caption{\label{tab:cu2o} Experimental results of the NQR
frequency $^{63}\nu_\text{Q}$ of cuprous oxide at 4.2~K and 300~K. For the 
other temperatures, see Fig.~\ref{fig:vst}.}
%%\begin{ruledtabular}
\begin{tabular}{llc}
\hline
$P$& $^{63}\nu_\text{Q}$& FWHM\\
(GPa)& (MHz)& (kHz)\\
\hline
\\
\multicolumn{3}{c}{4.2~K}\\
\\
0& 26.822& 18\\
4.37& 28.412& 104\\
5.28& 28.774& 135\\
7.11& 29.350& 150\\
8.37& 29.793& 260\\
9.18& 30.063& 300\\
\hline
\\
\multicolumn{3}{c}{300~K}\\
\\
0& 25.977& 17\\
4.26& 27.410& 55\\
5.12& 27.686& 63\\
6.83& 28.134& 75\\
8.01& 28.444& 100\\
8.78& 28.605& 150\\
\hline
\end{tabular}
%%\end{ruledtabular}
\end{table}
The dependence of $V$ on $P$ can be expressed in terms of the isothermal bulk
modulus $B_0$ and its pressure derivative $B_1$ by Murnaghan equation
of state\cite{MurnaghanEOS}.
\begin{equation}
\frac{V(0)}{V(P)} =\left(1 + \frac{B_1}{B_0}P\right)^{{1}/{B_1}}. 
\end{equation}
We assume power laws between the above EFG parameters and $V$:
\begin{align}
^{63}\nu_\text{Q}^{(0)}(P) &= {^{63}\nu_\text{Q}^{(0)}(0)}
\left(\frac{V(0)}{V(P)}\right)^\alpha,\\ 
\Theta(P) &= \Theta(0) \left(\frac{V(0)}{V(P)}\right)^\beta,\\
\lambda(P) &= \lambda(0)
\left(\frac{V(0)}{V(P)}\right)^\gamma. 
\end{align}
We then fit the high-pressure experimental data with $\alpha,
\beta, \gamma$, and $B_1$ as the parameters. The value of $B_1$ is known as
5.7,\cite{WernerHPCu2O} but the adjustment of $B_1$ provides a better fitting
with the standard deviation of 25~kHz. The results are:
\begin{align}
B_0 &= 131~\text{GPa  (fixed\cite{WernerHPCu2O})},\\
B_1 &= 7.74(36),\\
\alpha &= 1.928(19),\\
\beta &= -15.08(227),\\
\label{eq:cu2ohpfit}
\gamma &= 2.89(27).
\end{align}
The value of $^{63}\nu_\text{Q}$ is thus expressed at any
$(P, T)$ up to 300~K and about 9~GPa. It is noted that the relation $\alpha = 1$
is expected for an ideal ionic lattice.  The obtained value $\alpha$ of about 2
suggests additional effects of pressure, for example, strong modification in
covalency and/or in Cu-O bond length.

\begin{figure}[htbp]
\centering
\includegraphics[width=1.0\linewidth]{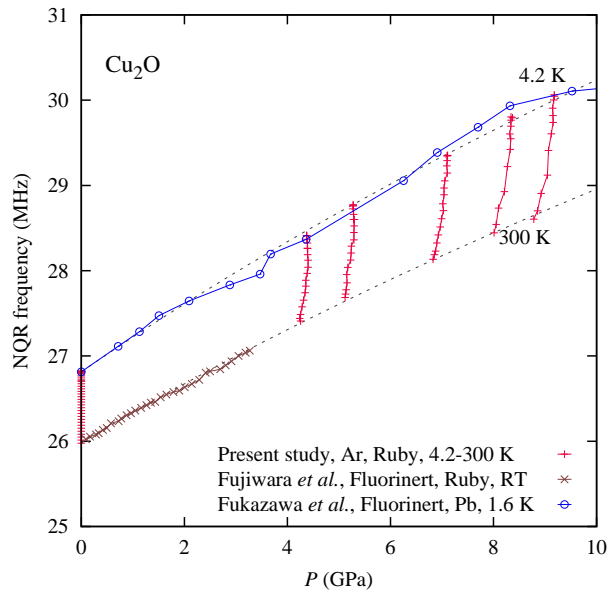}
\caption{(Color online) The $^{63}$Cu-NQR frequency in Cu$_{2}$O 
as a function of $P$. The red vertical lines show the present results, obtained by 
$T$-controlled runs after each clamp between 4.2~K and 300~K. 
Also plotted are the previous
studies\cite{FujiwaraNiCrAlcell,FukazawaBridgeman}.
The dashed lines are the isothermal curves given by the fitting formula at 4.2~K and at 300~K.}
\label{fig:cu2ocomp}
\end{figure}

Next, we compare our data with the previous results by 
Fujiwara~\textit{et\,al.}\cite{FujiwaraNiCrAlcell}, who used a piston 
cylinder cell, and by Fukazawa~\textit{et\,al.}\cite{FukazawaBridgeman},
who used a modified Bridgeman-type cell. The former was measured 
against the ruby scale at room temperatures.  In the latter experiments, 
pressure was determined at low temperatures from the superconducting 
transition temperature $T_\text{c}$ of lead. Note that the variation of
$T_\text{c}$ of lead has been studied by the ruby scale as
well.\cite{ThomassonPbTc} All the data points are potted in Fig.~\ref{fig:cu2ocomp}
as a function of $P$ to see the isothermal behavior. The curves (dashed lines in
Fig.~\ref{fig:cu2ocomp} at 4.2~K and at 300~K) obtained by
Eqs.~\eqref{eq:cu2o}--\eqref{eq:cu2ohpfit} reproduce all the experimental
results well within the errors.

The results demonstrate that the pressure dependence of
$^{63}\nu_\text{Q}$ in Cu$_{2}$O is well reproducible and explained by the simple
phenomenological formulations. The FWHM are very sharp, 20~kHz at ambient
pressure or 150~kHz at 9~GPa and 300~K, which correspond to 0.06, 0.43~GPa
of the $P$ variation respectively. Hence the cuprous oxide NQR
scale has been established against the ruby scale up to 9~GPa within at most 5\%
of errors.
Due to non-linear pressure dependence in ultrahigh-pressure 
range, the Newton method or its analogue must be used to obtain
pressure from the observed frequency. Alternatively, we suggest to look up 
Table~\ref{tab:conv} showing the calculated results of
Eqs.~\eqref{eq:cu2o}--\eqref{eq:cu2ohpfit}.
\begingroup
%%\squeezetable
\begin{table*}[p]
\caption{\label{tab:conv} The conversion table between presssure and the NQR
freuquency of cuprous or the Knight shift scales. These values are calculated
based on the fitted results and equations in text.}
%%\begin{ruledtabular}
{\tiny
\begin{tabular}{lcccccccc}
\hline
$P$& \multicolumn{2}{c}{Cu$_2$O: $^{63}\nu_\text{Q}$}&
\multicolumn{2}{c}{Pt$-$Cu: ${^{195}K} - {^{63}K}$}&
\multicolumn{2}{c}{$\beta$-Sn$-$Cu: ${^{119}K_\perp} - {^{63}K}$}&
\multicolumn{2}{c}{Pt$-$$\beta$-Sn: ${^{195}K} - {^{119}K_\perp}$}\\
(GPa)&
\multicolumn{2}{c}{(MHz)}& \multicolumn{2}{c}{(\%)}& \multicolumn{2}{c}{(\%)}&
\multicolumn{2}{c}{(\%)}\\
& 4.2~K& 300~K& 4.2~K& 300~K& 4.2~K& 300~K& 4.2~K& 300~K\\
\hline
0& 26.814& 25.980& -4.198&  -3.719& 0.437& 0.463& -4.635& -4.181\\
0.2& 26.897& 26.052& -4.193&  -3.716& 0.435& 0.460& -4.628& -4.175\\
0.4& 26.979& 26.124& -4.189&  -3.713& 0.432& 0.457& -4.620& -4.169\\
0.6& 27.060& 26.195& -4.184&  -3.710& 0.429& 0.453& -4.613& -4.164\\
0.8& 27.141& 26.265& -4.179&  -3.707& 0.426& 0.450& -4.606& -4.158\\
1& 27.221& 26.334& -4.175&  -3.705& 0.424& 0.448& -4.598& -4.152\\
1.2& 27.300& 26.403& -4.170&  -3.702& 0.421& 0.445& -4.591& -4.146\\
1.4& 27.378& 26.472& -4.166&  -3.699& 0.418& 0.442& -4.584& -4.141\\
1.6& 27.456& 26.539& -4.161&  -3.696& 0.416& 0.439& -4.577& -4.135\\
1.8& 27.533& 26.606& -4.157&  -3.694& 0.413& 0.436& -4.570& -4.130\\
2& 27.610& 26.673& -4.152&  -3.691& 0.411& 0.433& -4.563& -4.124\\
2.2& 27.685& 26.739& -4.148&  -3.688& 0.409& 0.431& -4.556& -4.119\\
2.4& 27.761& 26.804& -4.143&  -3.685& 0.406& 0.428& -4.550& -4.114\\
2.6& 27.835& 26.869& -4.139&  -3.683& 0.404& 0.426& -4.543& -4.108\\
2.8& 27.909& 26.933& -4.135&  -3.680& 0.402& 0.423& -4.536& -4.103\\
3& 27.982& 26.997& -4.130&  -3.677& 0.399& 0.421& -4.529& -4.098\\
3.2& 28.055& 27.060& -4.126&  -3.675& 0.397& 0.418& -4.523& -4.093\\
3.4& 28.127& 27.123& -4.121&  -3.672& 0.395& 0.416& -4.516& -4.088\\
3.6& 28.199& 27.185& -4.117&  -3.669& 0.393& 0.413& -4.510& -4.083\\
3.8& 28.270& 27.247& -4.113&  -3.667& 0.391& 0.411& -4.503& -4.078\\
4& 28.340& 27.308& -4.109&  -3.664& 0.389& 0.409& -4.497& -4.073\\
4.2& 28.410& 27.369& -4.104&  -3.661& 0.387& 0.406& -4.491& -4.068\\
4.4& 28.479& 27.429& -4.100&  -3.659& 0.384& 0.404& -4.484& -4.063\\
4.6& 28.548& 27.489& -4.096&  -3.656& 0.382& 0.402& -4.478& -4.058\\
4.8& 28.616& 27.549& -4.092&  -3.654& 0.381& 0.400& -4.472& -4.053\\
5& 28.684& 27.608& -4.087&  -3.651& 0.379& 0.398& -4.466& -4.049\\
5.2& 28.752& 27.666& -4.083&  -3.648& 0.377& 0.396& -4.460& -4.044\\
5.4& 28.819& 27.724& -4.079&  -3.646& 0.375& 0.393& -4.454& -4.039\\
5.6& 28.885& 27.782& -4.075&  -3.643& 0.373& 0.391& -4.448& -4.035\\
5.8& 28.951& 27.839& -4.071&  -3.641& 0.371& 0.389& -4.442& -4.030\\
6& 29.016& 27.896& -4.067&  -3.638& 0.369& 0.387& -4.436& -4.026\\
6.2& 29.081& 27.953& -4.063&  -3.636& 0.368& 0.385& -4.430& -4.021\\
6.4& 29.146& 28.009& -4.058&  -3.633& 0.366& 0.384& -4.424& -4.017\\
6.6& 29.210& 28.065& -4.054&  -3.631& 0.364& 0.382& -4.418& -4.012\\
6.8& 29.274& 28.120& -4.050&  -3.628& 0.362& 0.380& -4.413& -4.008\\
7& 29.337& 28.175& -4.046&  -3.626& 0.361& 0.378& -4.407& -4.003\\
7.2& 29.400& 28.229& -4.042&  -3.623& 0.359& 0.376& -4.401& -3.999\\
7.4& 29.462& 28.284& -4.038&  -3.621& 0.357& 0.374& -4.396& -3.995\\
7.6& 29.524& 28.337& -4.034&  -3.618& 0.356& 0.372& -4.390& -3.990\\
7.8& 29.586& 28.391& -4.030&  -3.616& 0.354& 0.371& -4.384& -3.986\\
8& 29.647& 28.444& -4.026&  -3.613& 0.353& 0.369& -4.379& -3.982\\
8.2& 29.708& 28.497& -4.022&  -3.611& 0.351& 0.367& -4.373& -3.978\\
8.4& 29.769& 28.549& -4.019&  -3.608& 0.349& 0.366& -4.368& -3.974\\
8.6& 29.829& 28.601& -4.015&  -3.606& 0.348& 0.364& -4.362& -3.970\\
8.8& 29.888& 28.653& -4.011&  -3.603& 0.346& 0.362& -4.357& -3.965\\
9& 29.948& 28.705& -4.007&  -3.601& 0.345& 0.361& -4.352& -3.961\\
9.2& 30.007& 28.756& -4.003&  -3.598& 0.343& 0.359& -4.346& -3.957\\
9.4& 30.065& 28.807& -3.999&  -3.596& 0.342& 0.357& -4.341& -3.953\\
9.6& 30.124& 28.857& -3.995& -3.995& 0.356& & -4.336& \\
9.8& 30.182& 28.908& -3.992& -3.992& 0.354& & -4.331& \\
10& 30.239& 28.957& -3.988& -3.988& 0.353& & -4.325&\\
\hline
\end{tabular}
}
%%\end{ruledtabular}
\end{table*}
\endgroup

\begin{figure}[htbp]
\centering
\includegraphics[width=1.0\linewidth]{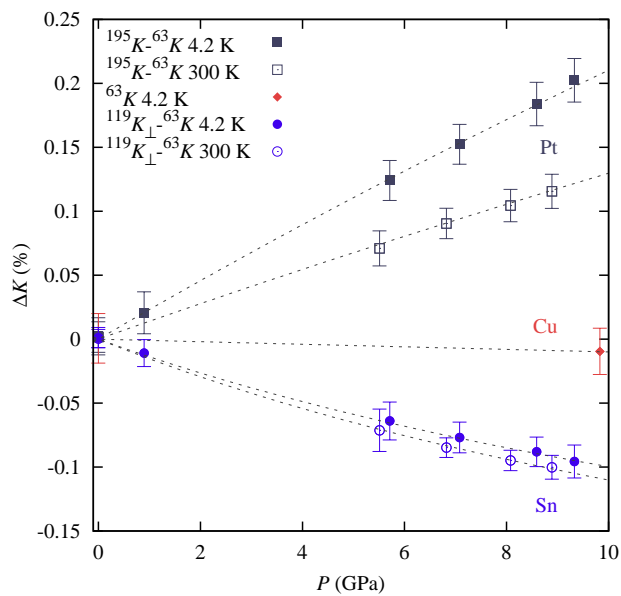}
\caption{(Color online) The variation of the Knight shifts, $\Delta K$, against
$P$. The dashed lines show the results of fitting (see text). For tin and
platinum, the shifts are determined against those of copper.}
\label{fig:dkvsp}
\end{figure}
\subsection{The Knight shift of $^{63}\text{Cu}$}
At ambient pressure, the Knight shift of copper metal is known to be almost
independent of $T$ and has a relatively small value of 0.20\%. 
These features make copper metal a favorable marker for field calibration 
at low temperatures, where a liquid NMR marker with sharp resonance line 
is not available. We determined the shift $^{63}K$ of Cu metal by comparing 
the resonance frequency of $^{63}$Cu and that of protons in glycerol.  
At room temperature glycerol shows a sharp line with the FWHM of 6~kHz 
shifted by 4~ppm relative to the signal in tetramethylsilane. 
Although it broadens to 16~kHz at 4.2~K, we assume the
chemical shift of glycerol is unchanged. At 300~K and 4.2~K, we
measured the shifts of copper at ambient pressure:
\begin{equation}
^{63}K = 
\begin{cases}
&0.2003(86)~\text{\% at 300~K}\\
&0.201(19)~\text{\% at 4.2~K}
\end{cases}.
\end{equation}
Here the gyromagnetic ratio of $^{63}$Cu nuclei is chosen to the value of the
IUPAC recommendation, 11.28933~MHz/T.\cite{IUPACNMR2008}
These values are consistent with the known values\cite{MetallicShifts} by
taking the same gyromagnetic ratio.
The number in the parenthesis
represents the calculated FWHM obtained by combining
the FWHM of the $^{63}$Cu and glycerol spectra independently.

Next, we determined the Knight shift of Cu metal under high pressure by using glycerol 
as the pressure transmitting medium.  At 9.8~GPa and 4.2~K, we obtained
\begin{equation}
^{63}K = 
0.191(18)\text{\%}.
\end{equation}
Then, we estimate the $P$-dependence of $^{63}K$ as,
\begin{equation}
\frac{\partial {^{63}K} }{ \partial P} = -10~\text{ppm/GPa}.  
\end{equation}
(See Fig.~\ref{fig:dkvsp}.) This small negative slope agrees well 
with the previous study at room temperature up to 1~GPa by 
Benedek~\textit{et\,al.}.\cite{BenedekPdep} Since copper is hard
material ($B_0 \simeq 140$~GPa), we consider that the following equation is
accurate enough to determine the field values in practical use for $P < 10$~GPa and $T \le 300$~K.
\begin{equation}
^{63}K(P, T) = 0.200 - 0.001P\text{ (\%)}.
\end{equation}

\subsection{The Knight shift of $^{195}\text{Pt}$}
Platinum is a well-known example of a paramagnetic metal in the 
vicinity of ferromagnetic instability. Unlike the ordinary metals, the magnetic 
susceptibility $\chi$ of Pt shows a strong $T$-dependence and is particularly 
enhanced at low temperatures.  The Knight shift of Pt metal $^{195}K$  
is generally composed of the spin part and the chemical shift. Since the 
former is proportional to $\chi$, $^{195}K$ is also largely enhanced
at low temperatures, $^{195}K \sim -4$\% and is highly $T$-dependent.
To use $^{195}K$ as a manometer, proper account is required for the $T$-dependence.

The self-consistent renormalization (SCR) theory provide a good framework to
describe magnetic susceptibility of paramagnetic metals close to magnetic instabilities.\cite{SCRBook}
For the low-$T$ limit, the SCR equation for the paramagnetic state near
ferromagnetic instability is analytically solved by approximating the digamma
function [$\log(u) - 1/2u - \phi(u) \approx u^{-2}/12$ for $u \gg 1$].
\begin{equation}\label{eq:scrapprox}
\chi^{-1} - \chi^{-1}_0 \propto T^2,
\end{equation}
where $\chi^{-1}_0 (> 0)$ represents the zero-temperature susceptibility.
In practice, the exponent of 2 may be adjusted for a better fit in high
temperatures. By assuming the same $P$-dependence both for the spin part and
chemical part of the Knight shift, $^{195}K$ is expected to have the following form:
\begin{align}\label{eq:ptshift}
K(P, T) &= K(P, 0) \left\{
 \frac{\sigma}{{1 +
(T/T_0(P))^\gamma}} + (1 - \sigma)\right\},\\
K(P, 0) &= K(0, 0)\left(\frac{V(0)}{V(P)}\right)^\alpha,\\
T_0(P) &= T_0(0) \left(\frac{V(0)}{V(P)}\right)^\beta.
\end{align}
Because the measurement of  $^{195}K$ relies on the $^{63}K$ field marker as
stated in the previous section, we hereby use the difference $^{195}K - ^{63}K$
as a manometer for better accuracy. 
\begin{equation}
^{195}K - ^{63}K =\left( \frac{^{195}f_\text{res}\times
11.28933}{^{63}f_\text{res}\times 9.152554} -1 \right)\times (1+{^{63}K}),
\end{equation}
where ${^{195}f_\text{res}}$ and ${^{63}f_\text{res}}$ are the resonant
frequencies of $^{195}$Pt and $^{63}$Cu nuclei respectively in the same
external field, and 9.152554~MHz/T is the gyromagnetic ratio of
$^{195}$Pt.\cite{IUPACNMR2008}
\begin{table}[h]
\caption{\label{tab:snpt} Experimental results of the Knight shifts of platinum
and tin at 4.2~K and 300~K. For the other temperatures, see
Fig.~\ref{fig:vst}.}
%%\begin{ruledtabular}
\begin{tabular}{lccc}
\hline
$P$& {${^{195}K} - {^{63}K}$}&
{${^{119}K_\perp} - {^{63}K}$}&
{${^{195}K} - {^{119}K_\perp}$}\\
(GPa)& (\%)& (\%)& (\%)\\
\hline
\\
\multicolumn{4}{c}{4.2~K}\\
\\
0& -4.196& 0.439& -4.634\\
0.897& -4.177& 0.427& -4.604\\
5.71& -4.074& 0.373& -4.447\\
7.08& -4.045& 0.361& -4.406\\
8.59& -4.014& 0.349& -4.363\\
9.33& -3.995& 0.342& -4.337\\ 
\hline
\\
\multicolumn{4}{c}{300~K}\\
\\
0& -3.717& 0.463& -4.180\\ 
5.51& -3.647& 0.392& -4.039\\
6.82& -3.628&  0.378& -4.006\\
8.08& -3.614& 0.368& -3.982\\
8.89& -3.603& 0.363& -3.965\\
\hline
\end{tabular}
%%\end{ruledtabular}
\end{table}

The results at several pressures and temperatures between 4.2~K and 300~K are
shown in Figs.~\ref{fig:vst} and \ref{fig:dkvsp} and in Table~\ref{tab:snpt}.
The data for ambient pressure are in very good agreement with the literature values\cite{MetallicShifts} 
after careful correction for the selection of field markers.
At ambient pressure, we obtain the parameters with the standard
deviation of 0.0014\%:
\begin{align}
({^{195}K} - {^{63}K})(0,0) &= -4.19804(55)\text{\%},\\
\gamma &= 2.750(38),\\
\sigma &= 0.273(10),\\
T_0(0) &= 338.2(81)~\text{ K}.
\end{align}

The equation of states of platinum has been studied extensively so that we
fixed $B_0$ and $B_1$ to the values given in the literatures\cite{ZhaEOSPt}. 
Then the values of $\alpha$ and $\beta$ are obtained as 
\begin{align}
B_0 &= 273.5~\text{GPa (fixed\cite{ZhaEOSPt})},\\
B_1 &= 4.70~\text{ (fixed\cite{ZhaEOSPt})},\\
\alpha &= -1.5230(43).\\
\beta &= 2.348(51).
\end{align}

We obtained the slope  $d^{195}K/dP$ in the low-$P$ limit to be 130~ppm/GPa at
300~K, and 220~ppm/GPa at 4.2~K. The value at 300~K is
consistent with the result reported by Matzkanin~\textit{et\,al.} up to 1.2~GPa.\cite{MatzkaninSnPbPt}

\subsection{The Knight shift of $^{119}\text{Sn}$}
The $\beta$-Sn metal is a typical Pauli paramagnet with week 
$T$-dependence of the susceptibility. Figure~\ref{fig:spectra}(c) shows 
$^{119}$Sn-NMR spectra of $\beta$-Sn.  Since $\beta$-Sn
has a tetragonal crystal structure, the spectrum exhibits a typical powder
pattern for anisotropic Knight shifts. We noticed that the shape of the entire spectrum 
depends on the sample form, most likely due to orientation of domains, 
which is especially prominent in foils. However, we found that the peak position, which 
corresponds to the in-plane Knight shift $K_\perp$, is independent of sample form.  
Therefore the in-plane shift $^{119}K_\perp$ is chosen as a manometer. By
measuring the spectrum of tin and copper at the same field, we determine 
\begin{equation}
{^{119}K_\perp} - {^{63}K} =\left( \frac{^{119}f_{\text{res}, \perp}\times
11.28933}{^{63}f_\text{res}\times 15.87700} -1 \right)\times (1+{^{63}K}).
\end{equation}

We fit the $T$-dependence of the difference of the Knight shifts 
${^{119}K_\perp} - {^{63}K}$ to Eq.~\eqref{eq:ptshift} at 
ambient pressure and obtained the following with the standard deviation of 0.00039\%,
\begin{align}
({^{119}K_\perp} - {^{63}K})(0,0) &= 0.43736(20)\text{\%},\\
\gamma &= 1.78(13),\\
\sigma &= -0.152(33),\\
T_0(0) &= 392(81)~\text{ K}.
\end{align}

For the ($P$--$T$)-dependence,
\begin{align}
B_0 &= 54.92~\text{GPa (fixed\cite{VaidyaCompMetals})},\\
B_1 &= 3.29(26),\\
\alpha &= -1.814(25),\\
\beta &= 1.59(15).
\end{align}
The standard deviation is 0.00088\%. The elastic constant $B_{0}$ is fixed to
54.92~GPa according to Ref.~\citen{VaidyaCompMetals}. Our value of  
$B_{1}$ is close to value in Ref.~\citen{VaidyaCompMetals} (3.651). For
the low-$P$ limit, the pressure derivatives of $^{119}K_\perp$ are
$-160$~ppm/GPa at 300~K, and $-150$~ppm/GPa at 4.2~K.
Note that ${^{119}K_\perp} - {^{63}K}$ exhibits non-linear
$P$-dependence due to softness of the material as shown in Fig.~\ref{fig:dkvsp},  in contrast to that of
platinum.
 
It is noteworthy that the $P$ dependence of the ${^{195}K} - {^{63}K}$ 
is opposite to that of ${^{119}K_\perp} - {^{63}K}$. Therefore, when one
observes the two scales to obtain the difference ${^{195}K} - {^{119}K_\perp}$,
more accurate $P$ determination can be expected. Another advantage of taking
the difference is to eliminate the errors arsing from measurement for the
$^{63}$Cu signals. See Table~\ref{tab:conv} in order to convert ${^{195}K}
- {^{119}K_\perp}$ to $P$ for simplicity.

\section{Conclusion}
In summary, we have established a new type of ultrahigh-pressure cell, by which we can
clamp pressure as high as 9~GPa. It was designed to achieve large sample space 
desired for microscopic experiments. As an example, we have measured the
pressure-temperature dependence of the NQR/NMR properties of several materials,
the NQR frequencies of cuprous oxide, and the Knight shifts of copper, platinum,
and tin against the ruby scale. These are well suited to determine pressure
during NQR/NMR experiments in situ.  We have presented fitting function to 
these sets of data, from which one can determine pressure up to 9~GPa at 
any temperature below 300~K.  The NQR frequencies of cuprous oxides has  
excellent resolution in determining pressure. NMR of copper in the wire can be used 
to determine the applied field.  The Knight shifts of platinum and tin are suitable 
manometers when inclusion of cuprous oxide in the sample coil is to be avoided. 

\section*{Acknowledgments}
We thank H.~Fukazawa for discussions and comments and J.~Yamazaki (ISSP
machine shop) for manufacturing the clamp cells. K.\,K. is financially supported
as a JSPS research fellow.

%\` %Just because of unusual number of tables stacked at end

\bibliography{64045} % Produces the bibliography via BibTeX.

%%\section*{Appendix}
%%\appendix

\end{document}